\documentclass[preprint,showpacs,preprintnumbers,amsmath,amssymb,floats]{revtex4}
\usepackage[dvips]{graphicx}
\usepackage[english]{babel}
\usepackage{amsmath}
\usepackage{amssymb}
\usepackage{color}
\usepackage{epstopdf}

\newcommand{\be}{\begin{eqnarray}}
\newcommand{\ee}{\end{eqnarray}}

\def\gsim{\mathop {\vtop {\ialign {##\crcr 
$\hfil \displaystyle {>}\hfil $\crcr \noalign {\kern1pt \nointerlineskip } 
$\,\sim$ \crcr \noalign {\kern1pt}}}}\limits}
\def\lsim{\mathop {\vtop {\ialign {##\crcr 
$\hfil \displaystyle {<}\hfil $\crcr \noalign {\kern1pt \nointerlineskip } 
$\,\,\sim$ \crcr \noalign {\kern1pt}}}}\limits}

\begin{document}
\noindent
\title{Renormalizations in unconventional superconducting states \\
born of normal and singular Fermi-liquids}

\author{Kazumasa Miyake$^{1}$ and Chandra M. Varma$^{2}$ }
\affiliation{
$^1$Center for Advanced High Magnetic Field Science, Osaka University, Toyonaka, 
Osaka 560-0831, Japan 
\\
$^2$Department of Physics and Astronomy, University of California, Riverside CA 92521, USA}

\pacs{}
\date\today
\begin{abstract}
The density of low energy particle-hole excitations is non-analytic in a singular Fermi-liquid, but it is altered on entering a superconducting state in which,
in the pure limit, it vanishes asymptotically at the chemical potential and in general is analytic. The single-particle excitations in the superconducting states are then quasi-particles so that a form of Landau theory may be constructed for thermodynamic and transport properties in the superconducting state.  In this theory, the renormalization of measurable properties due to quasi-particle interactions, such as specific heat, compressibility, magnetic susceptibility, superfluid density, etc. changes in a temperature dependent fashion from the non-interacting theory.  This is illustrated by showing the renormalization of these quantities and the relation between the parameters introduced to account for their temperature dependence. When the renormalizations in the normal state are large or singular, temperature dependence of properties in the superconducting states are then in general not useful for identifying the nodal character or symmetry of the superconducting state except for measurements at very low temperatures, upper limits of which are specified. The results obtained are expected to be useful in interpreting the experimental results for the temperature dependence of various properties in the superconducting state born of singular Fermi liquids. 
\end{abstract}
\maketitle

\section{Introduction}

The quasi-particle concept of Landau theory allows an understanding of thermodynamic properties and response functions in a Fermi-liquid. Its extension to the superconducting state allows an understanding of renormalization of properties in the superconducting state \cite{Larkin64, Leggett65}. A microscopic basis of the theory is the separation of the single-particle Green's function $G({\bf k}, \omega)$ into a coherent (or quasi-particle) part with poles and an incoherent analytic part. The theory is at its most powerful for calculation of the correlation functions of conserved quantities for which the incoherent part is shown to give no contribution. Then the perturbative calculations in terms of the quasi-particles and with analytic vertices or Landau parameters give controlled results in powers of $\omega/E_f$ and (${\bf k-k}_F)/k_F$. One can remain innocent of the subtleties while getting correct results. In the quantum-critical fluctuation regime of metals, where $G({\bf k}, \omega)$ have brach cuts, this innocence cannot be maintained. Only calculations using the specific low energy singularities of $G({\bf k}, \omega)$ {\it and} of the vertices can give correct results in such singular Fermi-liquids. Fortunately, this is possible for  some forms of quantum-criticality \cite{CMV2016}.

Although Landau theory cannot be applied to the singular Fermi-liquids \cite{CMV_Lorentz}, a form of Landau theory may still be applied to the superconducting state born of such liquids, because the density of low energy excitations in the superconducting state vanishes (in the pure limit) and together with that any low energy particle-hole singularities. The principles and methods for determining such renormalizations is discussed here, with application to the specific heat $C_v(T)$ and the London penetration depth or the superfluid density $\rho_S(T)$. Other properties may be calculated based on the same principles. There have been some excellent experimental results for specific heat and superfluid density \cite{Hashimoto2012, Bozovic2016, Armitage2017} in several superconductors whose normal state show evidence for a singular Fermi-liquid state. The interpretation of the results is expected to be aided by this paper.

Leggett \cite{Leggett65,Leggett75} has called superconductors born of a Fermi-liquid as Fermi-liquid superconductors (FL-superconductors). One may refer to superconductors born of singular Fermi-liquids as singular Fermi-liquid superconductors (SFL-superconductors).

The real (as opposed to virtual) scattering of fermions by the singular fluctuations also decreases the superconducting transition temperature \cite{MSV}. The absence of such fluctuations for $T \to 0$ implies that the ratio of the zero temperature gap to the transition temperature is enhanced above the BCS value \cite{MSV}. The temperature dependence of the gap is then also quite different from that in BCS \cite{LV}. This as well as the interactions between the quasi-particles in the superconducting state leads to temperature dependence in thermodynamic and transport properties which are quite different from the theory with non-interacting quasi-particles in the superconducting state.

In this paper, we also discuss the contrast between $C_v(T)$, which depends only on the thermally accessible density of states of excitations which give real scattering, and the ground state superfluid density $\rho_s(0)$, which depends on integrals over the complete frequency dependence of the conductivity. The temperature dependence of $\rho_s(T)$ however is related to the real scattering and therefore to the same parameters as the temperature dependence of $C_v(T)$. 

This paper is organized as follows.
First the region close to $T_c$ is discussed to show that the opening up of superconducting gap serves to cut-off the quantum-critical singularities and that their entropy is released through the enhancement of the specific heat at the superconducting transition.  Then the renormalizations in the specific heat, the compressibility and the magnetic susceptibility are considered followed by the renormalizations in the zero-temperature superfluid density and in its temperature dependent part.  We give the results also for Galilean Fermi-liquid superconductors obtained by Leggett \cite{Leggett65}, heavy Fermi-liquid superconductors in which the renormalizations appear quite differently than in a Galilean invariant superconductor \cite{CMV85}, and the singular Fermi-liquid superconductors.

\section{Fate of Singular Fermi-liquid effects in going from the normal to the superconducting state}

The necessary decrease of low energy particle-hole excitations in the superconducting state eliminates the singularity associated with the putative quantum-critical point (QCP) of the parent singular Fermi-liquid.
 Consequences for this on the thermodynamic properties just below $T_c$ may be estimated through the procedure in Ref. \cite{KosMV} adapted for a  general singularity in the quantum-critical response. 

The change of renormalizations just below $T_c$ for fluctuations of an order parameter $M({\bf r})$ can be studied adequately, if the superconducting transition has mean-field correlations, by a Ginzburg-Landau free-energy functional:
\be
\label{GLF}
{\cal F}(\Psi, M) =  V \Big[a(T_{c0}-T) |\Psi|^2 + b |\Psi|^4 + \frac{1}{V}\sum_{\bf k} \frac{1}{2}\Big[\chi^{-1}_{MM}({\bf k},T) + \lambda |\Psi|^2\Big]|M({\bf k})|^2 +...\Big].
\ee
The effect of biquadratic couplings between order parameters on the phase diagrams are of-course well known and have been considered for the AFM-superconducting couplings \cite{Schmalian2010}. The effect of the quasi-classical critical fluctuations at finite T in the free-energy has been specified by the static value of the real part of the temperature dependent susceptibility $\chi_{MM}({\bf k},T)$ of $M$. The effect of the high energy virtual fluctuations in promoting superconductivity and in direct coupling to $M^2$ is assumed to be taken into account in the coefficients $a, b, \lambda$ and in $T_{c0}$.


Following Ref.\cite{KosMV}, on integrating over the fluctuations of $M$ in the free-energy, one obtains the effective free-energy for superconductivity for dimension $d$,
\be
\label{effF}
{\cal F}_{sc}(\Psi) = a^*(T_c-T)|\Psi|^2 + b^* |\Psi|^4, \\
a^*(T-T_c) = a(T-T_{c0}) + \frac{d}{2} T \tilde{\chi}(T), 
\\ b^* = b - T^2 \frac{d}{2} \tilde{\chi}^2(T) \\
 \tilde{\chi}(T) = \sum_{\bf k} \lambda \chi({\bf k}, T).
\ee
These results are derived with the assumption that the change in free-energy due to the coupling $\lambda$ is small compared to the value for $\lambda =0$. 

The critical fluctuations in $M$ just below $T_c$ are also altered so that the equivalent susceptibility in the superconducting state is,
\be
\label{renchi}
\chi^{-1}_{MMs}({\bf k}, T) =  \chi^{-1}({\bf k}, T)+\lambda |\Psi|^2(T). 
\ee

The parameter $\lambda >0$ because the thermal fluctuations of $M$ are always pair-breaking. Therefore the $M^2|\Psi|^2$ coupling suppresses the region of the ordered phase of $M$ as well as the superconducting transition temperature. The quantum-critical point of $M$ is displaced to inside the region of order in the absence of superconductivity. 

From Eq. (\ref{effF}), one finds that the superconducting transition temperature is depressed from the mean-field value $T_{c0}$ to
\be
\label{Tc}
T_c \approx \frac{T_{c0}}{1+ \lambda \tilde{\chi(T_{c0})}}, 
\ee 
and that the mean-field specific heat jump at the transition increases by
\be
\label{leadspht}
\frac{\Delta \gamma^*}{\Delta \gamma} =
\frac{(a^*)^2/b^*}{ (a)^2/b}\approx 1 + \lambda T_c\tilde{\chi}(T_{c0}) + 2 \lambda^2 T^2_c\tilde{\chi}^2(T_{c0}).
\ee
This was used earlier \cite{KosMV} to understand the large jump in the specific heat at the superconducting transition in CeCoIn$_5$ using a particular form of $\chi({\bf k}, T)$, which is not quantum critical. Evidence is now available that this crystal is in fact close to quantum criticality. The large jump appears also in cuprates as well as in superconductors in the vicinity of an AFM quantum critical point \cite{Shibauchi2014, Carrington2013}. For the case of the topological fluctuations \cite{Aji-V-qcf1}, \cite{Hou-CMV-RG} which give rise to the marginal fermi-liquid \cite{MFL}, the critical fluctuations are product of functions of $q$ and of  $\omega/T$. The static long-wave-length susceptibility $\tilde{\chi}(T)$ then is proportional to $\log (\omega_c/T)$, where $\omega_c$ is the ultra-violet cut-off of the fluctuations. 

The results above are only valid for $T$ close to $T_c$.  The {\it details} of the low temperature properties,  since they depend on the low energy excitations in the superconducting state, cannot be obtained from such considerations. However, the enhanced jump in the specific heat at $T_c$ also gives necessary information on the low temperature thermodynamics through the requirement of entropy conservation and analyticity in the temperature dependence of properties, as we shall see below. 
The decrease of $T_c$ in Eq. (\ref{Tc}) is due to real or quasi-thermal scattering of the critical fluctuations. Similar results have been derived from more detailed considerations earlier \cite{MSV}.

 While there still exists a QCP in the superconducting state, the singularities in the properties of fermions at low energy must be strongly diminished for reasons given above. This may not be true in the unlikely case where quasi-static fluctuations of $M$ might induce a gap-less superconducting state with a decreased transition temperature  \cite{TVR-CMV}. This is unlikely because the effect of such fluctuations is expected to be overcome by the effective pairing interaction through the virtual fluctuations in $M$. We do not consider this subtle but unlikely situation here. 
  
 The diminution of the thermally accessible critical fluctuations in the superconducting state implies that the $T \to 0$ superconducting gap $\Delta(0)$ is not much changed from that in their absence. This means that the ratio of $\Delta(0)/T_c$ is enhanced with respect to the BCS value, as well as that the temperature dependence of the gap \cite{LV}  $\Delta(T)$ is quite different from BCS. This as well as interaction between superconducting quasi-particles changes the temperature dependence of physical quantities from that in a theory which ignores these considerations. 
 
 
 On the other side of the coin, the order parameter correlations $\tilde{\chi}({\bf k}, \omega, T)$ must show singularities of a different kind from the continuation of those above $T_c$ because the effect of coupling to the fluctuations of fermions, in particular the dissipation, changes. The $T \to 0$ order parameter correlation singularities are likely to be akin to the singularities in correlations  of a similar order parameter in an insulator.

\section{Thermodynamic Properties}

\subsection{Renomalization of Specific heat}

We will consider the renormalization of the specific heat at some length; The renormalization of the compressibility and the magnetic susceptibility follows from the same development. For non-interacting excitations in a superconductor, the specific heat is given in terms of the excitation energy \cite{Leggett75}
\be
\label{en}
 E_{\bf k}(T) = \sqrt{(\epsilon_{\bf k}-\mu)^2 + \Delta^2_{\bf k}(T)}
\ee
by
 \be
 \label{free-cv}
 C_{v0}(T) = - 2 \nu_{N}\int \frac{d{\hat{k}}}{4 \pi} \int_{E> \Delta({\bf k, T})} dE ~ 
 \frac{E^2}{\sqrt{E^2-\Delta^2({\bf k}, T)}}~\Big( \frac{E}{T} - \frac{dE}{dT}\Big)~\frac{d f(E)}{dE},
  \ee
where $f(E)$ is the Fermi-distribution function, and  $\nu_N$ is the density of single-particle excitations in the normal state,
$\nu_{N} = m^*k_F/\pi^2.$

First consider a FL-superconductor. Here we must distinguish the case of Galilean Fermi-liquids, as in the original Landau theory and its variants in lattices, especially the so called ``heavy Fermi-liquids" where the renormalization of physical properties by Landau parameters is quite different \cite{CMV85}. 

In the former, for example liquid $^3$He, the leading low temperature dependence of the specific heat does not {\it explicitly} \cite{Pines-Noz} depend on the Landau parameters either in its normal state or in  its superfluid state.
The specific heat coefficient $\gamma = C_v(T)/T$ is given in terms of the density of states or $1/|{\bf v}_F|$, which is defined as a ground state property. In Landau theory  for a Galilean invariant Fermi-liquid, no questions are permitted to be asked about the ground state. (However, using  Galilean invariance, it is shown that the current carrying velocity of quasi-particle excitations  is renormalized by $m/m^* = (1+F_1^s/3)^{-1}$.)  In the superconducting state, the same $1/|v_F|$ must enter in Eq. (\ref{free-cv}). 

 On the other hand, for heavy fermi-liquids, in which the assumption that the single-particle self-energy $\Sigma({\bf k}, \omega)$ is a weak function of momentum compared to energy is valid \cite{CMV85}, there is an enhancement of the specific heat by $ z^{-1} \approx (1+ F_0^s)$. Here $z$ is the quasi-particle renormalization amplitude (which cancels out against vertices, i.e. through Ward identities enforcing particle number conservation in low energy properties in a Galilean invariant Fermi-liquid, but not in a heavy Fermi-liquid) and $F_0^s$ is the s-wave spin symmetric Landau parameter. Higher $F's$ are much smaller than $F_0^s$, which is equivalent to the much weaker momentum dependence of $\Sigma({\bf k}, \omega)$ on ${\bf k}$ than on $\omega$. 
  
 Since $F_0^s$ depends on quasi-particle interactions, the specific heat renormalization depends on the density of excited quasi-particles through the density of states in the superconducting state and the temperature. Since the compressibility depends both on the specific heat mass as well as independently on $F_0^s$, it remains unrenormalized. 
 
In a singular Fermi-liquid, the leading temperature dependence of the specific heat depends on the density of states at the chemical potential through the velocity renormalization factor $z(T)z_k(T)$, where 
\be
z^{-1} = (1-\frac{\partial ({\rm Re}~\Sigma({\bf k}, \omega, T))}{\partial \omega})\Big|_{{\bf k} 
= {\bf k}_F, \omega = \mu}, ~~z_{\bf k} =1+{\bf v}_F^{-1}({\bf k})\cdot 
\frac{\partial ({\rm Re}~\Sigma({\bf k}, \omega, T))}{\partial {\bf k}}\Big|_{{\bf k} = {\bf k}_F, \omega = \mu}.
\ee
 For example, in a marginal Fermi-liquid $z^{-1}(T) = 1 + \lambda \log (\omega_c/\pi T). ~z_{\bf k} \approx 1$, because the self-energy again has negligible momentum dependence. If one is tuned close to criticality, such velocity or density of states renormalizations account for the logarithmic rise of the $C_v(T)/T$ as the temperature is decreased in the normal state. Such logarithmic enhancements of the entropy are directly observed in heavy-fermion \cite{HvLRMP2007} and cuprate quantum-criticality \cite{Taillefer-spht} and indirectly through measurements of thermopower in the Fe-based superconductors \cite{Carrington2013} near AFM quantum-criticality. 
 
For $T < T_c$,  the density of particle-hole excitations $\to 0$ at $\omega \to 0$. So such renormalization comes to a halt at $T< T_c$. However since this effect comes from the real part of the self-energy, which depends through the Kramers-Kronig relation on the absorptive part integrated over the complete frequency range, its value at $T_c$, i.e. $z^{-1}(T_c)$ continues essentially unchanged in the renormalization of the quasi-particle velocities of the superconducting state down to  $T \to 0$. The {\it leading} temperature dependence of the specific heat is then given by the density of states of the non-interacting particles of the superconducting state with the single particle density of states enhanced by
$z^{-1}(T_c)$.  This is true only with the assumption that $T_c$ is much less than the high frequency cut-off of the singular fluctuations.
 The factor $z^{-1}(T_c)$ multiplies the {\it leading} low temperature dependence of the specific heat $C_{v}(T)$, which is given by the density of states of the superconductor - exponential for a gapped state, $\propto T^2$ for a state with points of zeroes in two dimensions, etc.  
Therefore
  \be
 \label{spht -0}
Lim(T \to 0)  \frac{C_v(T)}{C_{v0}(T)} \approx  1,
\ee
 where $C_{v0}(t)$ gives the leading low temperature dependence of the specific heat.
 We take such renormalizations to include constant Fermi-liquid corrections in the normal state, which are also inevitably present beside the non-analytic terms. 
 
We must now consider the next order temperature dependent corrections due to interactions between excited quasi-particles in the superconducting states. Traditionally such corrections are avoided in Landau Fermi-liquid theory since the theory gives only the leading temperature dependent properties in $T/E_F$, to which there are no corrections if the density of excitations goes to 0 faster than $(T/E_F)^2$ 
as in s-wave type superconductors.  
We need them in the superconducting states if measurements are done  at temperatures comparable to the superconducting gap which sets the scale rather than the Fermi-energy so that higher powers of $T/\Delta(T)$ than the lowest need to be considered in order to compare with the experiments. The additional renormalizations come from the growth of renormalizations  as the density of thermally excited quasi-particles rises and from the temperature dependence of $\Delta(T)$. Except very close to $T_c$, the results in this paper are significant for experiments only for superconductors where the density of states of excitations is a power law and not for s-wave type superconductors. 

There are no general arguments for the magnitude of the renormalizations in the superconducting state. Approximate analytical calculations may be done.  More effective is to present an expansion of the {\it energy} in powers of the density of excitations and in terms of parameters to be determined. As explained below, knowing the results in the normal state puts constraints on the parameters.

Since the {\it energy} in the superconducting state is expected to be analytic as a function of temperature, a power series expansion for it in powers of the deviation of the distribution function from that at T=0 following Landau theory is permissible:
\be
\label{E}
\delta {\cal E} &= &\sum_{{\bf k},\sigma}  E_{{\bf k}}  \delta f_{\sigma}(E_{{\bf k}}) \nonumber \\
&+&
\frac{1}{2}\sum_{{\bf k},\sigma}\sum_{{\bf k}^{\prime},\sigma^{\prime}} 
g_{\sigma,\sigma^{\prime}}({\bf k},{\bf k'}) ~\delta f_{\sigma}(E_{{\bf k}}))~\delta f_{\sigma^{\prime}}(E_{{\bf k}^{\prime}}) +  O(\delta f)^3,
\ee
where  $\delta f_{\sigma}(E_{{\bf k}})$ is the deviation of the distribution function from that of the ground state,  and is given by the Fermi distribution 
$\delta f_{\sigma}(E_{{\bf k}})=1/[e^{\beta E_{{\bf k}}}+1]$ because $\lim_{T\to 0}\delta f_{\sigma}(E_{{\bf k}})=0$ 
except just at the nodes of $E_{\bf k}$ which give only constant contribution at most to the ground 
state energy. Note that the {\it chemical potential} for the quasiparticles in the superconducting state 
should be zero because the number of thermally excited quasiparticles is determined so as to 
minimize the free energy. 

Hereafter for considerations of the specific heat and for simplicity, we assume the interaction is contact type and 
spin-independent so that it is approximated as $g_{{\sigma,\sigma^{\prime}}}({{\bf k},{\bf k'}})\approx 
g_s$.  

To evaluate (\ref{E}), we need the density of states in the superconducting state which depends on its symmetry as well as several details. As a general form in the pure limit, applicable to gapless superconductors, we take the density of states measured from the chemical potential as defined by an exponent $\eta$ through
\be
\nu_S(E) = 
\sum_{{\bf k},\sigma}\delta(E-E_{\bf k})\approx A \nu_N |E|^{\eta}/\Delta^{\eta}.
\ee
where $A$ is a dimensionless constant of $O(1)$, and the approximate relation 
in the second equality is valid for $E\lsim \Delta$. 
With the use of the identity for arbitrary function $W(E)$ 
\be
\label{identity}
\sum_{{\bf k}\sigma}W(E_{\bf k})=\int_{0}^{\infty} dE\,W(E)\sum_{{\bf k}\sigma}\delta(E-E_{\bf k})
=\int_{0}^{\infty}dE\,\nu_{S}(E)W(E),
\ee
the specific heat from the first term in (\ref{E}) is
\be
\label{E1}
C_{v0}(T) = \frac{d}{dT}~\int_0^{\infty} dE \nu_{S}(E)
\frac{E}{e^{\beta E}+1}
&=& A_1(\eta) \nu_N \frac{T^{\eta +1}}{\Delta^{\eta}} 
(\eta+2-  \eta D_{\Delta/T} );\\
 D_{\Delta/T} &\equiv&\left|\frac{d \log \Delta}{d \log T}\right|;\\
A_1(\eta) &\equiv&  A(\eta +1)! \big(1-\frac{1}{2^{\eta +1}}\big) \zeta_R(\eta+2),
\ee
where $\zeta_R(y)$ is the Reimann zeta function. This term is the leading temperature dependence of the specific heat in which as argued by Eq. (\ref{spht -0}), the density of states or the effective mass is  renormalized approximately by $z^{-1}(T_c)$. Therefore in relation to non-interacting density of states $\nu_{N0}$, the interacting density of states is
\be
\nu_N \approx  z^{-1}(T_c) \nu_{N0}.
\ee
Apart from this renormalization (\ref{E1}) is identical to the non-interacting result of Eq. (\ref{free-cv}).
Then, 
$\delta n_{s0}(T)$, the density of un-renormalized excitations at a temperature $T$ is given by
\be
\label{delta_number}
\delta  n_{s0}(T) \equiv \int_0^{\infty} ~dE ~\nu_{S}(E)\frac{1}{e^{\beta E}+1} = A_2(\eta) \nu_N \frac{T^{\eta +1}}{\Delta^{\eta}};\\
A_2(\eta) = A \eta !(1-\frac{1}{2^{\eta}})\zeta_R(\eta+1).
\ee
The interaction term in (\ref{E}) is
\be
\label{E2}
\frac{1}{2}g_{s} \left[\delta n_{s0}(T)\right]^2.
\ee
The specific heat from the interaction term in (\ref{E}), $C_{v1}(T)$,  
is calculated by differentiating (\ref{E2}) with respect to $T$.  Also $C_{v0}(T)$ may be written as
\be
C_{v0}(T) =  \frac{A_1(\eta)}{A_2(\eta)}( 
\eta+2+\eta D_{\Delta/T})\,\delta n_{s0}(T),
\ee
assuming a gap monotonically decreasing as temperature is raised towards $T_c$

Adding $C_{v1}(T)$ to $C_{v0}(T)$ and dividing by T,

\be
\label{C1}
\frac{C_v(T)}{T} &=& \Big(1 +  g^s \left[\frac{A_2(\eta)}{A_1(\eta)}\right]^2\frac{\left(\eta+1+ {\eta}D_{\Delta/T}\right)}{\left(\eta+2+\eta D_{\Delta/T}\right)^2} \frac{C_{v0}(T)}{T} + \cdots  \Big)\frac{C_{v0}}{T}\\
\label{C2}
  &= &  \Big(1+ g^s \frac{A_2(\eta)}{A_1(\eta)}\frac{(\eta+1+ \eta D_{\Delta/T})}{(\eta+2+\eta D_{\Delta/T})} \frac{\delta n_{s0}(T)}{T} + \cdots \Big) \frac{C_{v0}}{T}.
\ee

 As temperature increases additional powers of the thermal excitation density $\delta n_{s0}(T)$ contribute which themselves are renormalized because of the interactions. The specific heat calculated for $T <T_c$ should match the value at $T_c$ and the total integrated entropy up to $T_c$ should equal the normal state entropy at $T_c$. Moreover the slope for $T$ just below $T_c$ should match the slope calculated from (\ref{leadspht}). These conditions can be met by using $\delta n_{s0}(T)$ as a temperature dependent mean-field function and writing the power series in Eqs. (\ref{C1}, \ref{C2}) as 
\be
\label{Cfinal}
\frac{C_v(T)}{T} \approx \Big(\frac{1}{1-g^s F(\eta, \Delta)\frac{C_{v0}}{T}} \Big)\frac{C_{v0}}{T},
\ee
where $F(\eta, \Delta) = \left[\frac{A_2(\eta)}{A_1(\eta)}\right]^2\frac{\left(\eta+1+ {\eta}D_{\Delta/T}\right )}{\left(\eta+2+\eta D_{\Delta/T}\right)^2} $. The form in (\ref{Cfinal}) is necessarily approximate since interactions involving higher powers of density will in general have different coefficients.

It is meaningful also to define
a temperature dependent quasi-particle renormalization amplitude $z_s(\eta, T)$ in the  superconducting state
\be
\label{finalC}
\frac{C_v(T)}{C_{v0}(T)} \simeq  z_s^{-1}(\eta,T). 
\ee
The definition and utility of $z_s(T)$ may be appreciated from the fact that $C_v(T)/T$ is proportional to the single-particle density of states, which in the approximation that the normal single-particle self-energy has negligible momentum dependence compared to its energy dependence is renormalized by a quasi-particle renormalization amplitude. 

 $C_v(T)$ must satisfy two conditions: $C_v(T_c)$ be equal to the value of the specific heat as $T$ approaches $T_c$ from below, and integral of $C_v(T)/T$  from 0 to $T$ must equal the integral of the equivalent quantity extrapolated from the normal 
state to $T \to 0$ containing the normal state renormalization amplitude $z^{-1}(T)$. The two parameters in Eq. (\ref{finalC}),  $g_s$ and $D_{\Delta/T}\equiv |d\log \Delta/d\log T|$ may be determined from these two conditions using the normal state values. Note that close below $T_c$, 
$D_{\Delta/T}$ may be a temperature dependent parameter. This is of-course not very satisfactory but inevitable given that one is looking here at the "high temperature" part of the superconducting state where no good expansions are possible.

Since $C_v(T)/T$ for fermions is generally related to the density of single-particle excitations, we may also write that a renormalized $\delta n^*(T)$ is related to the non-interacting $\delta n_{s0}(T)$ by
\be
\label{deltan}
\frac{\delta n^*_{s0}(T)}{\delta n_{s0}(T)} = z_s^{-1}(\eta,T).
\ee
In other words, the density of states $\nu_{s}(E)$ in Eq.\ (\ref{delta_number}) can be regarded as being 
renormalized by the same renormalization factor $z_{s}^{-1}(\eta,T)$ for the specific heat in 
Eq.\ (\ref{finalC}). 
This relation will be useful in the discussion on the superfluid density. 
 
 The results here imply that for strong or singular renormalization in the normal state, and the requirements that the renormalizations are strongly reduced in the superconducting state and must go over as $T \to T_c$ to the renormalizations in the normal state, it is in general incorrect to
deduce the symmetry of pairing from the power laws in various properties unless the leading $T/T_c << 1$ results are obtained in experiment and are unchanging in their temperature dependence over a credible range.

\subsection{Compressibility and Magnetic susceptibility}

As befitting specific heat, we have characterized $g({\bf k}, {\bf k'})$ by an s-wave parameter which includes both the contribution of the spin-symmetric and the spin-antisymmetric channels. Compressibility is essentially  un-renormalized 
in the case where strong local correlation exists as in heavy fermion metals, which is compatible to the fact that 
the self-energy is almost momentum independent \cite{CMV85}.  This implies that  the renormalization of the single-particle density of states, $z_s^{-1}(T)$, which determines the specific heat is exactly cancelled by a factor which may be called $[1+ F_{0s}(T)]$. 

For the magnetic susceptibility, the leading term and interaction terms in the coupling energy to magnetic field are functions of  $(\delta n_{{\bf p} \uparrow} - \delta n_{{\bf p} \uparrow})$. Then $g^a$, antisymmetric  spin coefficient should be used to get the change from the non-interacting values. Other than this change, the renormalization remain similar to that in Eq.\ (\ref{finalC}).

\section{Renormalization of Superfluid  density}
The zero frequency contribution to the conductivity of a superconductor,  $\frac{e^2}{m^{2}}\rho_S(T)\delta(\omega)$ defines the superfluid density $\rho_S(T)$. The text-book expression for the superfluid density for non- interacting electrons in the clean limit for a FL-superconductor,  i.e. for scattering rate $\tau^{-1} << \Delta_0$, is
\be
\label{rho}
\rho^{\alpha \beta}_S(T)  = \rho  \delta_{\alpha \beta} - \frac{m}{e^2} Lim_{q\to 0} Re~C^{\alpha \beta}_{jj}(q, 0; T),
\ee
 The density $\rho$ is the diamagnetic contribution which alone is present at zero temperature in a non-interacting system. The second term is the paramagnetic contribution given by the current-current correlation function $C^{\alpha \beta}_{jj}$ and is the normal fluid density $\big(\rho_N\big)^{\alpha \beta}(T)$, which is related to the current carried by thermal excitations. The London penetration depth $\Lambda_L$ is given by
 \be
\big(\Lambda_L^{-2}\big)^{\alpha \beta}(T) =  
\frac{4 \pi e^2}{m^{*2}c^2} \rho^{\alpha, \beta}_S(T).
\ee
 
The normal fluid density in the pure and non-interacting limits is,
\be
\label{FLrho}
\rho^{\alpha\beta}_{N0}(T) &=& \delta_{\alpha \beta}  \frac{m}{e^2} Re~C^{\alpha, \beta}_{jj}(q, 0, T) = m \hat{Y}(T),\\
Y^{\alpha, \beta}(T) &=&  6 \int \frac{d \hat{k}}{4\pi} \hat{k}^{\alpha} \hat{k}^{\beta} \int_{\Delta(\hat{k})}^{\infty}\frac{E}{\sqrt{E^2-\Delta^2({\bf k})}} \left[-\frac{df(E)}{dE}\right] dE \\
& \approx & \delta_{\alpha \beta} \frac{k_{\rm F}^{2}}{m} \int_0^{\infty} 
 dE\nu_{S}(E)\left[-\frac{df(E)}{dE}\right] \\
&=& \delta_{\alpha \beta} A_{\rho}(\eta)\frac{k_{\rm F}^2}{m} \nu_{N} \frac{T^{\eta}}{\Delta(T)^{\eta}};  \\
A_{\rho}(\eta) &=& A\eta!\left(1-\frac{1}{2^{\eta-1}}\right) \zeta_R(\eta)~ for ~\eta \geqslant 2, 
{\rm and } = \log (2)\,\,{\rm for }\,\, \eta =1. 
\ee
We now recall that the non-interacting density of states of quasiparticles $\nu_{N}$ may be written as $\nu_{N}=3\rho/k_{\rm  F}^{2}$. 
Therefore for non-interacting quasiparticles in the superconducting state is given as,
\be
\label{rho/n0}
\rho_{N0}(T) = 3\pi \rho A_{\rho}(\eta) \frac{T^{\eta}}{\Delta(T)^{\eta}}.
\ee
In order to discuss renormalizations for various models, with the use of Eq.\ (\ref{delta_number}),  we write the normal fluid density as
\be
\label{rho/n}
\rho_{N0}(T)=3\pi\rho\frac{A_{\rho}(\eta)}{A_{2}(\eta)}\frac{\delta n_{s0}(T)}{T\nu_{N}}.
\ee

In a Galilean invariant Fermi-liquid superconductor, there is no renormalization of the bare mass entering the superfluid density at $T=0$ because the current operator commutes with the Hamiltonian. The same result may be derived by satisfying the continuity equation. However the quasi-particle or normal fluid contribution has a back-flow correction to the current due to interaction between quasi-particles. Leggett has derived the correction to the normal fluid density in this case, such that the mass entering in the finite temperature reduction of ${\hat \rho}_s(T)$ is modified due to the Landau parameter $F_1^s$ \cite{Leggett65,Leggett75};
\be
\label{leggett}
{\hat \rho}_s(T) = \rho[{\hat 1}-{\hat Y}(T)]\left[{\hat 1}+\frac{F_{1}^{s}}{3}{\hat Y}(T)\right]^{-1}
\ee
This is effectively an interpolation between the bare mass at $T\to 0$ and the Fermi-liquid mass $m^* = m(1+F_{1}^{s}/3)$.

For heavy Fermi-liquids \cite{VMS1986} (or for electron-phonon interactions in the pure limit), i.e. Fermi-liquids in which the self-energy is a strong function of energy but a weak function of momentum \cite{CMV85}, the continuity equation is satisfied with a renormalization of the mass by $z^{-1} =(1+F_0^s)$.  
As a result, $m^{*}=m_{d}(1+F_{1}^{s}/3)$ with a dynamical mass $m_{d}\equiv mz^{-1}$ which  appears instead of $m$ in the formula [Eq.\ (\ref{leggett})] for the superfluid density tensor.

Now consider the effects of interactions between quasi-particles in the superconducting state on $\rho_N(T)$. Given Eq. (\ref{rho/n}) and Eq. (\ref{deltan}), the essential temperature dependence of $\rho_N(T)$ should be the same as of the density of single-particle excitations $\delta n(T)$ or $C_{v0}(T)/T$.  Also noted in Eq. (\ref{deltan}) is that, due to the interactions in the superconducting state, $\delta n_{s0}(T)$ is renormalized effectively to $\delta n_{s0}^*(T) = z_s^{-1}(T) \delta n_{s0}(T)$. Doing that leaves only the Leggett corrections due to $F_{1s}$. These may be included, but they are known to be quite small both in heavy fermions and in singular fermi-liquids compared to $z^{-1}$. 
We now recall, with the use of Eq.\ (\ref{leggett}), that the expression for ${\hat \rho}_{N}(T)$ is 
\be
\label{leggett2}
{\hat \rho}_{N}(T)=\rho(1+\frac{F_{1}^{s}}{3}){\hat Y}(T)\left[{\hat 1}+\frac{F_{1}^{s}}{3}{\hat Y}(T)\right]^{-1}
\approx \rho ~ {\hat Y}(T). 
\ee
We may now approximately include the effects of  higher powers of $\delta n_{s0}(T)$, as in the renormalization of the specific heat.  Ignoring the small difference between $A_{\rho}(\eta)$ and ${A_2(\eta)}$
\be
\label{rhon1}
\rho_N(T) \approx \rho_{N0}(T) \left[z_s(\eta,T)\right]^{-1},
\ee
where the renormalization factor $z_{s}(\eta,T)$ is given through Eqs.\ (\ref{Cfinal}), (\ref{finalC}) and (\ref{deltan}). 

The measured $\rho_N(T)$ and $C_v(T)$ can be used to test that the renormalization is by the same $z_s^{-1}$. 

\subsection{\it Zero Temperature}

 We now briefly discuss the zero temperature superfluid density, which has been a focus of some recent experiments \cite{Bozovic2016} \cite{Armitage2017} and calculations \cite{Hirschfeld2018}.

For superconductors with elastic or impurity scattering, including gapless superconductors, the paramagnetic term is not $0$ at $T \to 0$, and so the superfluid density is reduced in that limit. But there is an additional contribution in SFL-superconductors. The singularity in the self-energy in the normal state in cuprates leads to the normal state conductivity $\sigma_N(\omega,T)$, which is proportional to $1/\omega$ \cite{MFL}, \cite{CMV2016} which is cut-off at low energies by the impurity scattering rate and at high frequencies both by the logarithmic correction to the mass and the ultra-violet cut-off $\omega_c$ in the singular fluctuation spectra. (Ignoring the necessary high frequency cut-off in the predicted scale invariant optical conductivity \cite{MFL} has led to peculiar power law fits to the $\sigma(\omega)$ at high frequencies by some experimentalists and theorists). The part of the paramagnetic conductivity with $\omega >> 2 \Delta(T)$ persists in the superconducting state down to $T\to 0$. To be in accord with the sum-rule on the total conductivity, $\rho_S(T=0)$ is then reduced by about
\be
\frac{m^{*}}{ \frac{2}{\pi}e^{2}}\int_{\approx 2\Delta}^{\omega_c} d\omega ~\sigma_N(\omega).
\ee
This is closely related to $z_s^{-1}(T=0)$ introduced as a renormalization of the mass in the superconducting state at $T=0$.

Some discrepancies \cite{Bozovic2016,Armitage2017} were noted recently in the superfluid density in overdoped cuprate La$_{2-x}$Sr$_x$CuO$_4$ at various $x$.  Overdoped cuprates also have paramagnetic conductivity extending well above the superconducting gap, as befits a material with a cross-over from a singular Fermi-liquid to a Fermi-liquid but only for low enough frequencies.
The discrepancies appear to have been accounted for qualitatively by calculations of the effects of impurities \cite{Hirschfeld2018} in d-wave superconductors. We suggest that the considerations above, including the renormalization of the superfluid density may be useful in getting quantitative agreement with the experiments.

 \section{Summary and Conclusions}\label{CR}

\begin{figure}[tbh]
\centering
\includegraphics[width=1.0\columnwidth]{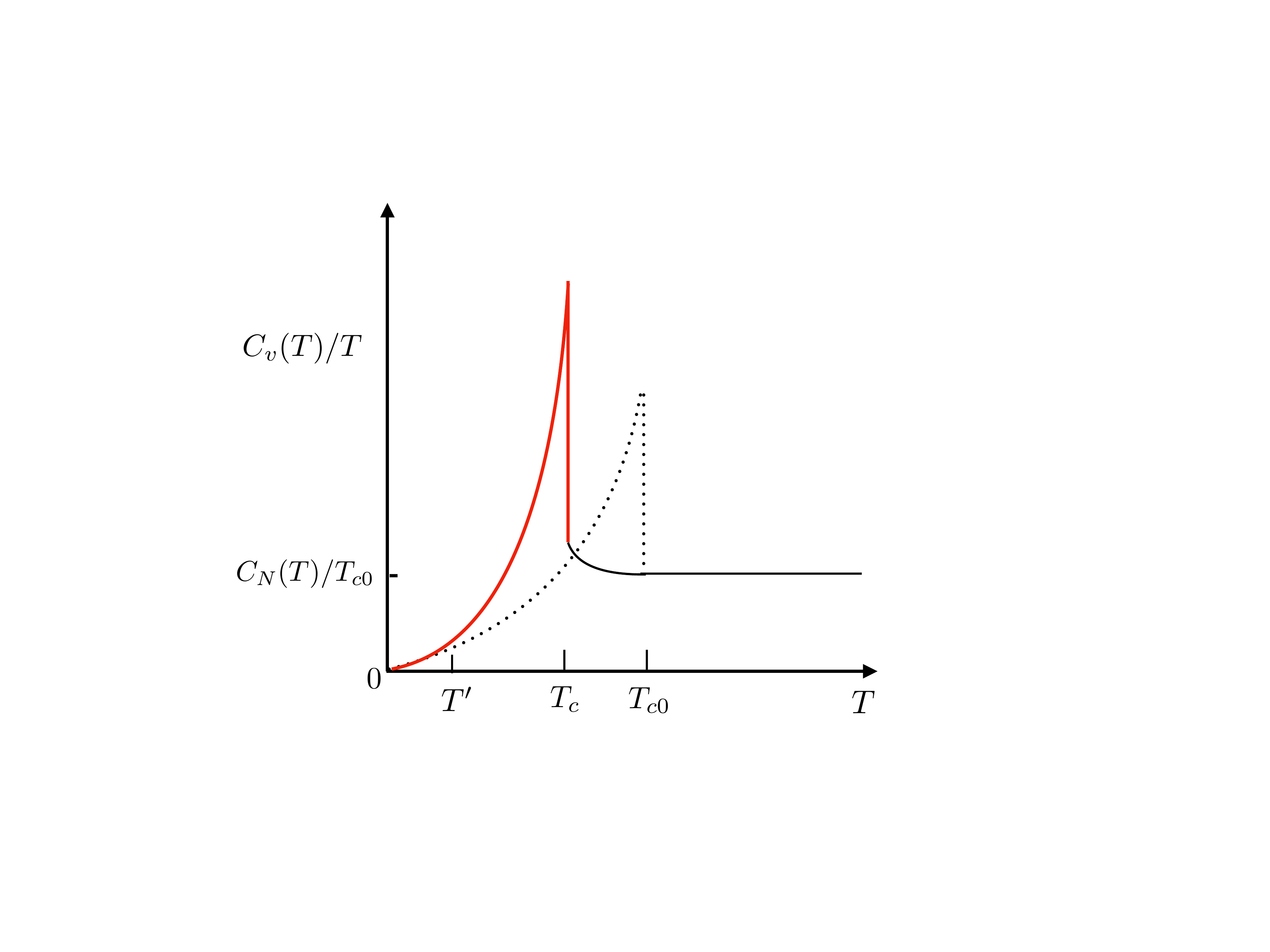}
\caption{Figure illustrating the renormalizations in the specific heat coefficient $C_v/T$, and by the arguments of this paper in other properties, in a heavy Fermion superconductor. The $T\to 0$ renormalization of this quantity is similar to the Fermi-liquid renormalization for $T \gtrsim T_c$. But due to the depression of $T_c$ due to the large inelastic scattering and the consequent high $\Delta C_v/T$ at the superconducting transition, the specific heat decreases rapidly below $T_c$ compared to the non-interacting case. The temperature dependence of physical quantities is unlike those of the non-interacting power laws unless measurements are made below $T'$, estimated in the paper.}
\label{Fig:1}
\end{figure}

\begin{figure}[tbh]
\centering
\includegraphics[width=1.0\columnwidth]{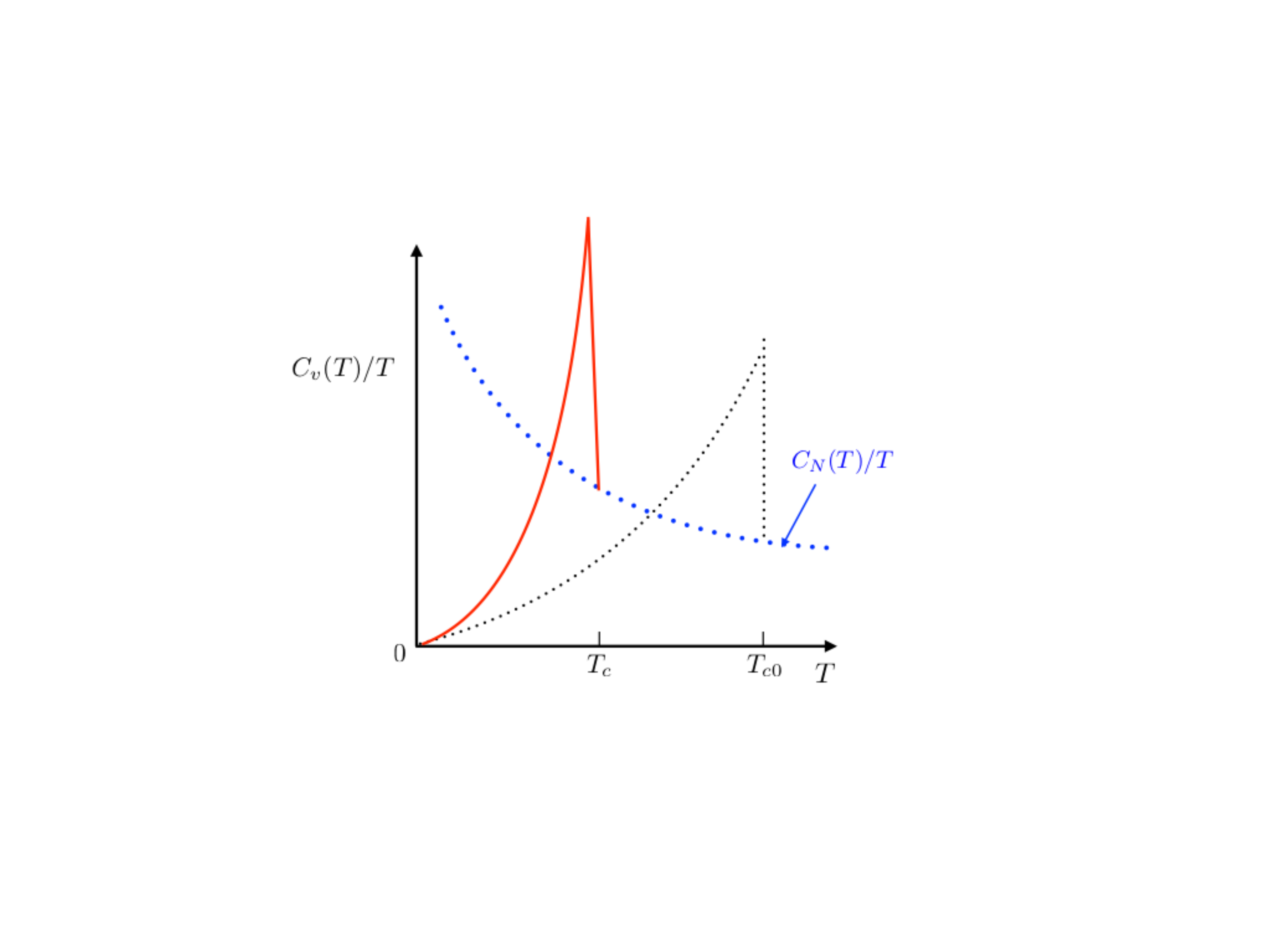}
\caption{Figure illustrating the renormalization in the specific heat coefficient $C_v/T$, and by the arguments of this paper in other properties, in a singular Fermi-liquid superconductor. Beside the considerations for heavy-Fermi-liquid superconductors, there is the additional fact that the continuation of the normal
$C_v/T$ to low temperatures has a $\log (\omega_c/T)$ singularity. The superconductor must then have additional temperature dependence to make up the sum-rule on the entropy integrated to $T_c$.}
\label{Fig:2}
\end{figure}

In this paper we have considered the renormalization of measurable properties such as specific heat, magnetic susceptibility,
compressibility and superfluid density in the superconducting states for metals (in the pure limit) which have large Fermi-liquid renormalizations
(heavy Fermi-liquids) in the normal state or those which are near quantum-critical points of an order parameter competing with superconductivity (singular Fermi-liquids). The analysis is restricted to situations in which the self-energy of the fermions is weakly momentum dependent compared to the frequency dependence. Many interesting metallic compounds being studied fall into this category (but liquid He-3 does not). For the case of singular Fermi-liquids, besides the cuprates, rare-earth compounds and the superconducting Fe based compounds in the vicinity of their antiferromagnetic instability also appear to belong to this category \cite{CMV2016}. The signature for this is their common linear in $T$ contribution to the low temperature resistivity in the normal state accompanied by a $T \log T$ specific heat whose magnitude is consistent with a real part of the self-energy $\propto \omega \log \frac{\omega_c}{\omega}$  coming from the entire Fermi-surface.

This paper was motivated by excellent experiments which were fitted to give power laws  over limited temperatures for various properties in the superconducting states
which are however hard to understand, for instance in Refs. \cite{Hashimoto2012}, \cite{Bozovic2016}, \cite{Armitage2017}. 
An essential result of this work is that fit to a power law should not be expected, except at very low temperatures, in the systems being discussed because of the necessarily changing renormalization with temperature in the superconducting state.
These changes are qualitatively and quantitatively  different for heavy Fermi-liquids (i.e. systems with a small  $z$ in the normal state) and singular Fermi-liquids in which $z \to 0$ logarithmically as temperature goes towards $0$. In the latter case, the cut-off in the logarithmic singularity in the superconducting state introduces new features not found in the former case. The effect of this persists down to quite low temperatures. In both cases, the decrease in $T_c$ due to inelastic scattering of fermions from fluctuations induces a large jump (in mean-field theory) on the specific heat at the superconducting transition. To balance the entropy, the specific heat decreases very rapidly below $T_c$.

The two cases are illustrated by schematic sketches of $C_v(T)/T$ in this section. In Fig. (\ref{Fig:1}), the heavy-Fermi liquid case is sketched. $T'$ denotes the upper limit of the temperature below which the power laws are those expected from the simple noninteracting quasi-particle theory. This is obtained from Eq. (\ref{Cfinal}).  At low temperatures, the $\Delta$ dependence of $F(\eta, \Delta)$ is unimportant.  $T'$ is estimated simply from $g_s C_{v0}(T')/T' \approx 1$. This gives
\be
\label{T'}
T^{\prime}\approx \Delta_{0}
\frac{\eta+2}{Ag_{s}\nu_{N}\eta !}
\left[
\frac{\zeta_{R}(\eta+2)}{\zeta_{R}(\eta+1)}\,\frac{2^{\eta+1}-1}{2^{\eta+1}-2}\right]
\ee
The factors depending on $\eta$ give only O(1) and are not so important as $g_s \nu_N$. The reduction of $T'/\Delta_0$ is inversely as $g_s \nu_N$. This does not account for the change from the non-interacting specific heat at higher temperatures due to the larger jump at $T_c$ due to the renormalizations and the reduction $T_c/T_{c0}$ given in Eqs. (\ref{leadspht}) and (\ref{Tc}). 

In Fig. (\ref{Fig:2}), a singular Fermi-liquid case is sketched. The normal state specific heat and its extrapolation below $T_c$ is shown as a dashed line.  In this case there is the additional consideration of the sum-rule on the entropy up to $T_c$  that the singular Fermi-liquid $C_v/T$ extrapolated to the superconducting state has a $\log (\omega_c/T)$ dependence, where $\omega_c$ is the upper cut-off in the singularity in the normal state fluctuation spectrum. This will be distributed from low temperatures all the way to $T_c$. In general then the low temperature limit of a pure power law is reduced further below the value given in (\ref{T'}).

To compare the detailed temperature dependence of the specific heat with the calculations, it is necessary to use Eq. (\ref{Cfinal}) and use the procedure given there to estimate the coefficients. But given the measurement of one property, other properties may be obtained more easily.
From Fig. (\ref{Fig:1}) and (\ref{Fig:2}), one may estimate $z_s^{-1}(T)$ using Eq. (\ref{finalC}), and estimate variation in the temperature dependence of  other properties, such as superfluid density or magnetic susceptibility, due to the changing renormalizations,  from the results given above. A recent comparison of the measured  renormalizations in specific heat and the normal fluid density \cite{Shu2018} are consistent with both being given by the same temperature dependent factor.


\end{document}